\documentclass{article}
\newcommand{\bi}{\begin{itemize}}
\newcommand{\ei}{\end{itemize}}
\newcommand{\bq}{\begin{quote}}
\newcommand{\eq}{\end{quote}}
\newcommand{\ket}[1]{\vert#1\rangle}

\newcommand{\braket}[2]{\langle#1\vert#2\rangle}
\newcommand{\ketbra}[2]{\vert#1\rangle\langle#2\vert}

\newcommand{\RRR}{$I\hspace{-0.25em}R^3$}
\begin{document}
\title{Objective probability and quantum fuzziness}
\author{U. Mohrhoff\\
Sri Aurobindo International Centre of Education\\
Pondicherry 605002 India\\
ujm@auromail.net}
\maketitle
\begin{abstract}This paper offers a critique of the Bayesian interpretation of quantum mechanics with particular focus on a paper by Caves, Fuchs, and Schack  containing a critique of the ``objective preparations view'' or OPV. It also aims to carry the discussion beyond the hardened positions of Bayesians and proponents of the OPV. Several claims made by Caves et al. are rebutted, including the claim that different pure states may legitimately be assigned to the same system at the same time, and the claim that the quantum nature of a preparation device cannot legitimately be ignored. Both Bayesians and proponents of the OPV regard the time dependence of a quantum state as the continuous dependence on time of an evolving state of some kind. This leads to a false dilemma: quantum states are either objective states of nature or subjective states of belief. In reality they are neither. The present paper views the aforesaid dependence as a dependence on the time of the measurement to whose possible outcomes the quantum state serves to assign probabilities. This makes it possible to recognize the full implications of the only testable feature of the theory, viz., the probabilities it assigns to measurement outcomes. Most important among these are the objective fuzziness of all relative positions and momenta and the consequent incomplete spatiotemporal differentiation of the physical world. The latter makes it possible to draw a clear distinction between the macroscopic and the microscopic. This in turn makes it possible to understand the special status of measurements in all standard formulations of the theory. Whereas Bayesians have written contemptuously about the ``folly'' of conjoining ``objective'' to ``probability,''  there are various reasons why quantum-mechanical probabilities can be considered objective, not least the fact that they are needed to quantify an objective fuzziness. But this cannot be appreciated without giving thought to the makeup of the world, which Bayesians refuse to do. Doing this on the basis of how quantum mechanics assigns probabilities, one finds that what constitutes the macroworld is a single Ultimate Reality, about which we know nothing, except that it manifests the macroworld or manifests itself as the macroworld. The so-called microworld is neither a world nor a part of any world but instead is instrumental in the manifestation of the macroworld. Quantum mechanics affords us a glimpse ``behind'' the manifested world, at stages in the process of manifestation, but it does not allow us to describe what lies ``behind'' the manifested world except in terms of the finished product---the manifested world, for without the manifested world there is nothing in whose terms we could describe its manifestation.
\end{abstract}

\section{Introduction}
\label{sec:intro}The present paper has two objectives. Fuchs~\cite{Fuchs2003} and Caves, Fuchs, and Schack~\cite{CFS2002a,CFS2002c} have presented a case for a Bayesian interpretation of quantum mechanics. My first objective is to offer a critique of the Bayes\-ian interpretation with particular focus on the authors' recent paper~\cite{CFS2007} (``CFS'' in what follows). CFS contrast their subjective stance on both quantum uncertainty \textit{and} certainty with what they call the ``objective-preparations view'' (OPV). According to the latter, a system's quantum state is determined by a sufficiently detailed, agent-independent classical description of a preparation device, this being itself an agent-independent physical system. My second objective is to carry the discussion beyond these hardened opposing positions. 

The first section states what is \textit{not} at issue. Points of divergence are spelled out in Sec.~\ref{sec:divergence}, the key issue being the nature of the time dependence of quantum states. CFS share with the proponents of the OPV the cognate notions that quantum states \textit{evolve} and that quantum-mechanical probabilities are \textit{absolute}. Here the case is made that quantum-mechanical probabilities are conditional rather than absolute, and that the time dependence of a quantum state is the dependence of an algorithm, which serves to assign probabilities to the possible outcomes of a measurement, on the time of the measurement. 

CFS go so far as to assert that for sufficiently divergent prior beliefs, ``two agents might even legitimately assign different pure states.'' This brings up the question, considered in Sec.~\ref{sec:differ}, of how much state assignments can legitimately differ. A further claim made by CFS is that proponents of the OPV ignore the essential quantum nature of preparation devices (and that it is wrong to do so). They illustrate this claim with the help of a quantum circuit containing qbits and c-\textsc{not} gates. The holes in their argument are pointed out in Sec.~\ref{sec:qbits}, and in Sec.~\ref{sec:device} it is argued that a device is a legitimate preparation device if and only if its quantum-mechanical nature is irrelevant to its functioning as a preparation device.

The present paper's constructive part begins with Sec.~\ref{sec:stdistinctions}. While quantum Bayesians keep the notorious measurement problem safely locked away in ``a single black-boxed piece of hardware: the 1-Qbit measurement gate''~\cite{Mermin2006}, most others believe that ``to solve this problem means to design an interpretation in which measurement processes are not different in principle from ordinary physical interactions,'' as an anonymous referee once put it to me.%
\footnote{Since quantum mechanics describes interactions in terms of correlations between the probabilities of the possible outcomes of \textit{measurements} performed on the interacting systems, I wonder what the referee could have meant by an ``ordinary physical interaction.''}
The way I see it, to solve this problem means to design an interpretation in which the central role played by measurements in standard axiomatization of quantum mechanics is recognized and explained. If this were easy, it would have been done long ago. As was pointed out by Dieks~\cite{Dieks}, the difficulty is that
\bq
the outcome of foundational work in the last couple of decades has been that interpretations which try to accommodate classical intuitions are impossible, on the grounds that theories that incorporate such intuitions necessarily lead to empirical predictions which are at variance with the quantum mechanical predictions. However, this is a negative result that only provides us with a starting-point for what really has to be done: something conceptually new has to be found, different from what we are familiar with. It is clear that this constructive task is a particularly difficult one, in which huge barriers (partly of a psychological nature) have to be overcome.
\eq
I believe that in attempting this constructive task, the following points, which are outlined in Sec.~\ref{sec:stdistinctions}, must be borne in mind~\cite{M2000,M2004,M2005,M2008}:
\bi
\item The intrinsically and completely differentiated arena of physical events called ``spacetime'' is a figment of our mathematical imagination. 
\item Spacetime coordinates only exist as features of the quantum-mechanical correlation laws. While not objective \textit{per se}, they are (to a limited extent) capable of being objectified or becoming objective via outcome-indicating events or states of affairs---the only points of contact that exist between the correlation laws and the physical world.%
\footnote{A measurement is an event or state of affairs from which something can be inferred about something else. (Attempted but unsuccessful measurements are not included in this definition.) Do we also need to define events and states of affairs? Obviously not. For one thing, we know them when we see them. For another, because quantum mechanics presupposes their existence or occurrence, it cannot account for it, anymore than we can explain why there is anything, rather than nothing at all.}
\item The spatiotemporal aspects of the physical world consist of the more or less \textit{fuzzy} relative positions and relative times that exist between physical objects, physical events, or physical states of affairs.
\item These relative positions and relative times exist to the extent that they are indicated by, or can be inferred from, physical objects, physical events, or physical states of affairs.
\item The spatiotemporal differentiation of the physical world is incomplete.
\ei
Outcome-indicating events and states of affairs occur or obtain in the macro\-world. What counts as ``macroscopic'' is rigorously defined (without going\break
beyond the conceptual tools of standard quantum mechanics) in Sec.~\ref{sec:macroworld}.  This definition, which strongly relies on the incompleteness of the world's spatiotemporal differentiation, makes it legitimate to attribute to the macroworld a reality independent of anything external to it.

How does one define and quantify a fuzzy observable? The answer, spelled out in Sec.~\ref{sec:probability}, is: by assigning probabilities to the possible outcomes of a measurement of the same. This is also the reason why the probabilities of measurement outcomes play a central role in standard axiomatization of the theory. And why are observables such as relative positions and relative momenta fuzzy? The answer to this question is that the fuzziness of those observables is required for the stability of matter. This answer disposes of arguments by CFS to the effect that if quantum uncertainty is subjective, then so is certainty. On the contrary, quantum uncertainty---a bad translation of Heisenberg's term \textit{Unsch\"are}, the literal meaning of which is ``fuzziness''---can be as objective as certainty. After all, the stability of a material object hinges on the objective fuzziness of its internal relative positions and momenta, rather than on our  subjective uncertainty about the values of these observables.

In Sec. \ref{sec:objective} several good reasons are given why quantum-mechanical probabilities may be considered objective. In addition, an argument by Fuchs and Schack~\cite{FuchsShack2004}, designed to illustrate ``the folly of trying to have two kinds of probabilities in quantum mechanics'' (one subjective and one objective), is refuted.

Perhaps the most disconcerting aspect of quantum mechanics---which only comes to light if, instead of trying to explain it away, one tries to make sense of the central role played by measurements---is the supervenience of the microscopic on the macroscopic. A set of properties~A supervenes on a set of properties~B if and only if any two objects which share all properties in~B must also share all properties in~A. The properties of the microworld depend in just this way on the properties of the macroworld. Hence we cannot think of particles, atoms, and such as constituents of the macroworld. Then what constitutes the macroworld? This question is addressed in Sec.~\ref{sec:UR}. The conclusion reached is that the number of ``ultimate constituents of matter'' is exactly one. This conclusion is supported by a time-honored ontology: ultimately there exists a One Being, and the macroworld is its manifestation. The so-called ``microworld'' is therefore neither a world nor a part of any world. Rather, it is instrumental in the manifestation of the macroworld. Quantum mechanics affords us a glimpse ``behind'' the manifested world at formless particles, non-visualizable atoms, and partly visualizable molecules which, instead of being the world's constituent parts or structures, are instrumental in its manifestation. But---and this is \textit{why} the microscopic supervenes on the macroscopic---we cannot describe what lies ``behind'' the manifested world except in terms of the finished product---the manifested world.

Section~\ref{sec:manifestation}, finally, spells out the reason why the manifestation of the world cannot be described except in terms of the manifested world. It is simply that without the manifested world there is nothing in whose terms we could describe its manifestation. Section \ref{sec:conclusion} summarizes and concludes.

\section{Points of agreement}
\label{sec:agreement}Before embarking on the critical part of this paper, I wish to point out what is \textit{not} at issue. I agree with CFS that 
\bi
\item[(1)]  there are no preassigned values to, and no instruction sets behind, quantum measurement outcomes; this includes outcomes that are certain;
\item[(2)]  differently put, there is nothing intrinsic to a quantum system---no element of reality, no independently possessed property---that guarantees a particular measurement outcome;
\item[(3)]  there is no local \textit{and} realistic explanation for the correlations predicted by quantum mechanics;
\item[(4)]  state assignments, including pure-state ones, have no objective status; 
\item[(5)]  the role of physical law in a quantum mechanical world is to relate probabilities, not to determine them;
\item[(6)]  the Bayesian interpretation of probability is superior to both the frequentist and the potentiality/propensity interpretations.%
\footnote{Relative frequencies, belonging as they do to the domain of facts or events, are different from probabilities, and the  interpretation of probabilities as potentialities~\cite{Heisenberg1958,Shimony1989} or propensities~\cite{Gillies1973,Suppes1973,Giere1979,Popper2000} contravenes the conditional (as against absolute) nature of quantum-mechanical probabilities pointed out in the next section.}
\ei

\section{Points of divergence}
\label{sec:divergence}A quantum state $\rho(t)$ is an algorithm that serves to assign probabilities to the possible outcomes of any measurement that may be made at the time~$t$. In keeping with this commonplace, I take the view that the time dependence of $\rho(t)$ is a dependence on the time of a measurement---the measurement to whose possible outcomes $\rho$ serves to assign probabilities. It is \textit{not} the time dependence of an evolving state of any kind. This disposes of the question whether a quantum state has two modes of evolution or only one; it has \textit{none}.%
\footnote{To forestall a possible objection: if the Hamiltonian depends on time, it includes effects of \textit{classical} boundary conditions. By no means does its possible time dependence support the view that quantum states evolve.}

What is at issue here is \textit{not} (at any rate, not primarily) the role of probability in quantum mechanics but \textit{the principle of evolution}. By this I mean the notion that physics can be divided into a kinematical part, which concerns the description of a system at an instant of time, and a dynamical part, which concerns the evolution of a system from earlier to later times. While relativity made it seem as if the principle of local action was a consequence of the principle of evolution, quantum mechanics appears to rule out not only local realistic interpretations of the predicted and observed correlations between measurements performed in spacelike relation but also realistic interpretations of the principle of evolution.

In keeping with this principle, the wave function is widely regarded as the primary affair, which the propagator $\braket{x_f,t_f}{x_i,t_i}$ serves to propagate through time \textit{\`a la}%
\footnote{In the case of a nonrelativistic system consisting of a fixed number~$n$ of particles, $x$~stands for a set of $3n$ coordinates.}%
\begin{equation}
\psi(x_f,t_f)=\int\!dx_i\,\braket{x_f,t_f}{x_i,t_i}\,\psi(x_i,t_i).
\end{equation}
This way of thinking makes it seem as if quantum-mechanical probabilities were determined by a wave function or a quantum state,  rather than by measurement outcomes via computational devices called ``wave functions'' or ``quantum states.'' If in addition one believes that quantum states exist and evolve in the absence of measurements, and that measurements merely \textit{contribute} to determine them, then one treats quantum-mechanical probabilities as \textit{absolute} probabilities. 

Conversely, the assumption that quantum-mechanical probabilities are absolute, leads to (i)~the spurious question of what it is that determines them (over and above measurement outcomes) and (ii)~the facile answer that they are determined by a wave function or a quantum state that is \textit{not} fully determined by measurement outcomes. This in turn leads to (iii)~the spurious question about the nature of quantum states and (iv)~the false dilemma that quantum states are \textit{either} states of nature \textit{or} states of knowledge or belief. One horn of this dilemma is the OPV, the other is the belief that quantum states exist and evolve in the absence of measurements \textit{as prior probabilities in the beliefs of agents}. While this notion can make it easier to accept the idea that measurements contribute to determine the evolving states of quantum systems, it is nevertheless unwarranted, inasmuch as it involves the belief in the existence of evolving quantum states.

To my mind, the laws of quantum mechanics encapsulate correlations between measurement outcomes---diachronic correlations between outcomes of measurements performed on the same system at different times as well as synchronic correlations between outcomes of measurements performed on different systems in spacelike relation. We can use these correlations to assign probabilities to possible measurement outcomes \textit{on the basis of actual measurement outcomes}, just as in a hypothetical classical world we can use the classical laws to predict later states on the basis of earlier ones. Our use of the correlations for the purpose of predicting probabilities may be considered subjective, but the correlations themselves are as objective as the classical laws are in a classical world.%
\footnote{I am not saying that a mathematical construct like the Faraday tensor $\mathbf{F}$ is anything more than a computational tool. I am only saying that $\mathbf{F}$ is a useful mathematical tool for formulating an objective correlation law. The essential difference I see between the laws of classical physics and those of quantum physics is the absence of principles limiting the accuracy of predictions in classical physics as against the presence of such principles in quantum physics.}

If the laws of quantum mechanics are correlation laws---correlating measurement outcomes synchronically as well as diachronically---then quantum-mechanical probabilities are conditional rather than absolute. Conversely, if quantum-mechanical probabilities are conditional, as has been stressed by Primas~\cite{Primas2003},%
\footnote{Primas~\cite{Primas2003} has also drawn attention to an axiomatic alternative to Kolmogorov's~\cite{Kolmo1950} formulation of probability theory, due to R\'enyi~\cite{Renyi1955,Renyi1970}. Whereas in Kolmogorov's theory absolute probabilities have primacy over conditional ones, R\'enyi's theory is based entirely on conditional probabilities. Primas states that every result of Kolmogorov's theory has a translation into R\'enyi's.}
then we have no need for an evolving quantum state, be it a state of nature or a state of belief. In the following passage CFS appear to support this conclusion:

\bq Imagine a scientist who performs a sequence of $Z$ measurements on a qbit. Quantum mechanics, plus his experience and prior judgment and perhaps the outcomes of a long sequence of previous measurements, make him certain that the outcomes will all be ``up.'' Now he performs the measurements, and he always gets the result ``up.'' Shouldn't the agent be surprised that he keeps getting the outcome ``up''? Doesn't this mean that it is a fact, rather than a mere belief, that the outcomes of his experiment will be ``up''?\dots\ The answer to the first question is easy: Surprised? To the contrary, he would bet his life on it. Since the agent was certain that he would get the outcome ``up" every time, he is not going to be surprised when that happens. Given his prior belief, only observing ``down'' would surprise him, since he was certain this would not happen, though nature might choose to surprise him anyway. The answer to the second question is similarly straightforward. According to our assumption, the agent has put together all his experience, prior beliefs, previous measurement outcomes, his knowledge of physics and in particular quantum theory, all to predict a run of ``up'' outcomes. Why would he want any further explanation? What could be added to his belief of certainty? He has consulted the world in every way he can to reach this belief; the world offers no further stamp of approval for his belief beyond all the factors that he has already considered.

\eq In other words, we do not need a state of affairs that is external to the agent, that depends on the first measurement outcome, and that is responsible for the outcomes of the subsequent measurements. We do not need an evolving state of affairs that \textit{mediates} an influence of the first measurement outcome on the subsequent measurement outcomes. In fact, if mediating states of affairs fail to account for the synchronic correlations, how can we expect them to account for the diachronic ones? 

Instead of drawing the obvious conclusion that quantum states encapsulate \textit{unmediated} correlations between measurement outcomes, CFS do not challenge the notions at the roots of the above dilemma. They accept the notion that quantum states evolve, for how else could they assert that a pre-measurement system-apparatus state arises from a ``unitary interaction'' between the system and the apparatus? They accept that quantum-mechanical probabilities are absolute rather than always conditional on measurement outcomes, for how else could they assert that ``facts alone never determine a quantum state''? They merely deny the external, agent-independent existence of quantum states. In other words, they merely internalize the evolving quantum states of their opponents. Whereas for proponents of the OPV a quantum state exists in advance of measurements as a state of nature, for CFS it exists in advance of measurements in the form of a prior belief. ``Any usage of probability theory starts from a \textit{prior probability assignment}'' (original emphasis). And whereas that state of nature is unknown, this prior belief can be neither verified nor falsified. ``The question of whether a prior probability assignment is true or false cannot be answered.'' The Bayesian internalization of quantum states may relieve symptoms of the disease but it is far from being a cure.

\section{How much can state assignments differ?}
\label{sec:differ}CFS go so far as to assert that for sufficiently divergent prior beliefs, ``two agents might even legitimately assign different pure states''.%
\footnote{Fuchs and Shack \cite{FuchsShack2004} write that ``there is no fact of nature to prohibit two different agents from using distinct pure states $\ket\psi$ and $\ket\phi$ for a single quantum system.'' Of course, if quantum states are only beliefs, they can be mutually inconsistent, as beliefs often are---even if they are held by the same person.}
This brings up the question of how much state assignments can legitimately differ. This question has been addressed by Brun \textit{et al.}~\cite{Brunetal2002}, who showed that several density matrices are mutually compatible if and only if the supports of all them have at least one state in common, or equivalently, if and only if all of them have expansions of the form 
\begin{equation}
\rho=\sum_i p_i\,\ketbra{v_i}{v_i},\qquad p_i>0
\end{equation}
with at least one state common to all expansions. It follows at once that two pure-state density matrices are compatible if and only if they are identical.

Brun \textit{et al.} define a set of density matrices to be compatible when there could be circumstances under which they would represent the knowledge different people have of one and the same physical system. Caves, Fuchs, and Schack~\cite{CFS2002b}, who call the belief that an outcome is impossible a ``firm belief,'' have shown that this definition of compatibility is equivalent to the existence of a density operator that does not contradict the firm beliefs of any party. Hence if two agents can ``legitimately assign different pure states,'' as CFS claim, such a density operator does not exist. \textit{Any} density operator then contradicts someone's belief that an outcome is impossible (i.e., it assigns a probability greater than~0 to at least one outcome to which at least one party assigns probability~0). 

\section{An argument based on qbits}
\label{sec:qbits}Carl Sagan, echoing Hume and Laplace, popularized the slogan that ``extraordinary claims require extraordinary proof.'' The argument offered by CFS in support of their extraordinary claim that two agents can legitimately assign different pure states, however, is not likely to convince anyone who is not already convinced of the conclusion. The argument begins by considering a quantum circuit containing a system qbit initially in the state $\alpha\ket0+\beta\ket1$ and an ``apparatus'' qbit initially in the state $\ket0$.  A controlled NOT (c-\textsc{not}) gate produces the entangled state $\alpha\ket{00}+\beta\ket{11}$. A ``measurement'' of the apparatus qbit then flips the system qbit if the outcome is~1, otherwise it does nothing. The system qbit is thus prepared in the state~$\ket0$.

CFS next consider a modified circuit in which the ``measurement device'' is replaced by a second c-\textsc{not} gate. This likewise prepares the system qbit in the state~$\ket0$ but leaves the apparatus qbit in the state $\alpha\ket0+\beta\ket1$. If this modified circuit is used but the initial state of the apparatus qbit is~$\ket1$, the system qbit is prepared in the state~$\ket1$, and the apparatus qbit is left in the state $\alpha\ket1+\beta\ket0$. The system qbit is thus prepared in the initial apparatus state, whatever that state has been.

CFS insist on ``the {\it essential\/} quantum nature of the preparation device'' (original emphasis). In conjuction with their Bayesian view of quantum states, this ``means that the prepared quantum state always depends on prior beliefs in the guise of a quantum operation that describes the preparation device.'' The operation of a preparation device ``always depends on prior beliefs about the device, in particular, its initial quantum state.''

In actual fact, even if the operation of a preparation device did sometimes depend on the initial state of the device, this won't always be the case. If, for instance, in the above example we simply exchange the roles of system and apparatus, then the final apparatus state is identical to the initial system state, whatever that state has been. Hence one might simply point out that the device chosen by CFS is not a legitimate preparation device. A preparation device is legitimate only if its initial state does \textit{not} affect the state prepared and, as we have just seen, such devices exist.

One might also object against the use of a two-state system as a measurement device. How will one ever know whether a failure of the device to change its state indicates an actual outcome or simply a failure of the device? (There is no such thing as a 100\% efficient device.) To get around this problem---and to forestall qbit-based shenanigans such as the above---a measurement device is usually expected to have a neutral state in addition to its outcome-indicating states. 

\section{The preparation device: classical or quantum?}
\label{sec:device}Another tenet of the view advocated by CFS  is that the assumption that a preparation device can be given a complete classical description ``neglects that any such device is quantum mechanical and thus cannot be specified completely in terms of classical facts.'' It may or may not be the case that a preparation device can be given a completely classical description, but it is besides the point, inasmuch as a device is a legitimate preparation device only if its quantum-mechanical nature is irrelevant to its functioning as a preparation device.

As an illustration, consider a particle of spin-$\frac12$ that has been prepared, with the help of a Stern-Gerlach filter, in the state $\ket{+z}$. For a zero Hamiltonian this means that a subsequent measurement of the particle's spin component with respect to the $z'$~axis, performed with a Stern-Gerlach splitter and two detectors, will yield ``up'' or ``down'' with  probabilities%
\footnote{If we take into account that no detector is 100\% efficient, these probabilities are the respective limits of $n_u/n$ and $n_d/n$ as $n\rightarrow\infty$, where $n$ is the total number of detector ``clicks'' and $n_u$ and $n_d$ are the respective numbers  of ``up'' and ``down'' clicks.}
$|\braket{+z'}{+z}|^2$ and $|\braket{-z'}{+z}|^2$, respectively. In order to say this, neither the preparation device nor the measurement device needs to be given any but a classical description. The essential piece of information is the angle between the respective directions of two magnetic field gradients. All we need to know about the detectors is that a response (``click'') indicates the arrival of a particle in its sensitive region. The prior beliefs of agents nowhere enter the picture.

All that quantum mechanics allows us to calculate is correlations between measurement outcomes. To be able to make predictions (or retrodictions, for that matter), all we need and all we have is (i)~these correlations and (ii)~actual measurement outcomes. Think of density operators as machines with inputs and outputs: insert a measurement M by specifying its possible outcomes, insert the time of~M, insert at least one relevant actual outcome of another measurement, press \textsc{compute}---and out pop the probabilities of the possible outcomes of~M. The \textit{only} possible description of a quantum system---at any rate, the only one that enters into the calculation of probabilities and can therefore be tested---consists of quantum-mechanical correlation laws and actual measurement outcomes, the former given in mathematical language, the latter given in \textit{classical} terms. If there is anything more, it will involve hidden variables and/or a cryptodeterminism of some kind. Speculations involving such things invariably raise questions that they fail to answer---e.g., ``Why are Bohmian trajectories unobservable?''---even though there are straightforward answers---e.g., ``because they don't exist.'' 

Come to think of it, why is it that, according to CFS, the question of whether a prior probability assignment is true or false, cannot be answered? The answer is that without factual input, the quantum-mechanical correlation laws are useless. If you press \textsc{compute} without inserting at least one measurement outcome, the probabilities that pop out will evince an equal lack of information, in keeping with the Principle of Indifference.

\section{The limited reality of spatiotemporal\\
distinctions}
\label{sec:stdistinctions}The disease alluded to at the end of Sec.~\ref{sec:divergence} is the notion that a time coordinate serves to label objectively distinct simultaneities---infinitely thin slices of spacetime, if you like. In other words, it is the belief that physical events and states of affairs happen or obtain in an intrinsically and completely differentiated arena called ``spacetime.'' The principle of evolution mentioned in that section is a direct consequence of this belief, inasmuch as it underwrites the notion that quantum states are evolving instantaneous states. 

What if that intrinsically and completely differentiated arena is a figment of our mathematical imagination? In this case spacetime coordinates only exist as features of the quantum-mechanical correlation laws. While not objective \textit{per se}, they are (to a limited extent) capable of being objectified or becoming objective via outcome-indicating events or states of affairs---the only points of contact that exist between the correlation laws and the physical world. The spatiotemporal aspects of the physical world then consist of the more or less \textit{fuzzy} relative positions and relative times that exist between physical objects, physical events, and physical states of affairs. And these relative positions and relative times exist to the extent that they are indicated by, or can be inferred from physical objects, physical events, or physical states of affairs.%
\footnote{This means, \textit{inter alia}, that if physical objects, events, and states of affairs did not exist, neither would the spacetime ``arena.'' The existence of the world's spatiotemporal properties stands or falls with the existence of objects, events, or states of affairs to which they can be attributed.}

If I am on the right track, then the spatiotemporal differentiation of reality doesn't go ``all the way down.'' To see this, remember that the exact localization of a particle implies an infinite momentum dispersion and thus an infinite mean energy. (In a relativistic world, the attempt to produce a strictly localized particle instead results in the creation of particle-antiparticle pairs.) Therefore no material object ever has a sharp position (relative to any other object). Now let \RRR($O$) be the set of exact positions relative to some object~$O$. We can conceive of a partition of \RRR($O$) into finite regions that are so small that none of them is the sensitive region of an actually existing detector.%
\footnote{By a \textit{detector} I mean an object with a sensitive region~$R$, capable of indicating the presence in~$R$ of another object of a more or less specific type. An object~$A$ is capable of \textit{indicating} the possession of a property (by another object) or a value (by an observable) if this can be inferred from a property---or a change in the properties---of~$A$.}
Hence we can conceive of a partition of \RRR($O$) into sufficiently small but finite regions $R_k$ of which the following is true: there is no object~$Q$ and no region $R_k$ such that the proposition ``$Q$~is inside~$R_k$'' has a truth value. In other words, there is no object~$Q$ and no region $R_k$ such that $R_k$ exists for~$Q$. But a region of space that does not exist for any material object, does not exist at all. The regions $R_k$---or, what comes to the same, the distinctions we make between them---correspond to nothing in the physical world. They exist solely in our heads. It follows that the spatial differentiation of the physical world is incomplete---it doesn't go all the way down. 

What holds for the world's spatial differentiation also holds for its temporal differentiation. The times at which observables possess values, like the values themselves, must be indicated in order to exist or be possessed. Clocks are needed not only to indicate time but also, and in the first place, to make times available for attribution to indicated values. Since clocks indicate times by the positions of their hands,%
\footnote{Digital clocks indicate times by transitions from one reading to another, without hands. The uncertainty principle for energy and time, however, implies that such a transition cannot occur at an exact time, except in the unphysical limit of infinite mean energy~\cite{Hilgevoord1998}.}
the world's incomplete spatial differentiation implies its incomplete temporal differentiation.

\section{The macroworld}
\label{sec:macroworld}The possibility of obtaining evidence of the departure of an object~$O$ from its classically predictable position%
\footnote{A ``classically predictable position'' is a position that can be predicted on the basis of (i)~a classical law of motion and (ii)~all relevant value-indicating events.}
calls for detectors whose position probability distributions are narrower than~$O$'s---detectors that can probe the region over which $O$'s fuzzy position extends. For most objects with sufficiently sharp positions, such detectors do not exist. For the objects I have in mind, the probability of obtaining evidence of departures from the classically predictable motion is very low. Hence \textit{among} these objects there will be many of which the following is true: every one of their indicated positions is consistent with every prediction that can be made on the basis of previously indicated properties and a classical law of motion. These are the objects that deserve to be labeled \textit{macroscopic}. To permit a macroscopic object---e.g., the proverbial pointer needle---to indicate the value of an observable, one exception has to be made: its position may change unpredictably if and when it serves to indicate a property or a value.

We are now in position to define the \textit{macroworld} unambiguously as the totality of relative positions obtaining between macroscopic objects. Let's shorten this to ``macroscopic positions.'' By definition, macroscopic positions never evince their fuzziness (in the only way they could, through departures from classically predicted values). This makes it legitimate to attribute to the macroworld---not individually to each macroscopic position but to the macroworld in its entirety, and not merely ``for all practical purposes'' but strictly---a reality independent of anything external to it (such as the consciousness of an observer~\cite{LonBau,vN,Wigner,Squires,Lockwood,Albert,Stapp}). But it also entails the supervenience of the microscopic on the macroscopic.

In philosophy, supervenience is a relationship of dependence typically obtaining between sets of properties. According to one standard definition, a set of properties~A supervenes on a set of properties~B if and only if any two objects which share all properties in~B must also share all properties in~A. Beyond that, the nature of this dependence (logical, nomological, or other) remains unspecified. A typical example is the notion that psychological properties supervene on physical properties (usually goings-on in a brain), which does not imply that the former depend on the latter logically or nomologically. What I mean by the supervenience of the microscopic on the macroscopic is that the properties of the microworld depend in just this way on the properties of the macroworld---rather than the other way round, as we are wont to think.%
\footnote{We may now think of measurements as departures of macroscopic positions from their classical law of motion, indicating goings-on in the microworld rather than being caused by them.}

\section{Probability, objective fuzziness, and\\
the stability of matter}
\label{sec:probability}The Bayesian approach to quantum semantics is to take the Bayesian interpretation of probability for granted and to see what light it throws on quantum mechanics. To my mind, this puts the cart before the horse. As Appleby~\cite{Appleby2005} remarked, ``[w]hereas the interpretation of quantum mechanics has only been puzzling us for about 75 years, the interpretation of probability has been doing so for more than 300 years.'' Quantum mechanics, properly understood, may be better equipped to throw light on the meaning of probability than probability theory is equipped to throw light on the interpretation of quantum mechanics. In any event, the Bayesian interpretation of probability appears to me to be singularly ill-equipped for this task.

One only has to ask: what accounts for the stability of matter? Specifically, what accounts for the existence of spatially extended objects 
\bi
\item that are composed of a large but finite number of unextended objects (particles that do not ``occupy'' space), 
\item that ``occupy'' finite volumes of space, and 
\item that neither explode nor collapse the moment they are formed?
\ei
The existence of such objects hinges on the \textit{objective fuzziness} of the relative positions and relative momenta of their constituents.%
\footnote{By itself, the fuzziness of a relative momentum causes the corresponding relative position to become more fuzzy. In a stable material object, this inherent tendency of relative positions to become more fuzzy is counterbalanced by the electrostatic attraction between oppositely charged particles, which causes relative positions to become less fuzzy.}
This, rather than our \textit{subjective uncertainty} about the values of these observables, is what ``fluffs out'' matter. 

The mathematical theory of probability relies on a primitive notion of probability to make contact with reality~\cite{Appleby2006}. Instead of treating probability as a primitive notion, we may regard possibility as primitive and look upon probability as a quantification of possibility. Any further interpretation ought to take account of the context in which probabilities are assigned. The one-size-fits-all approach fails to do justice to the disparate roles that probability plays in the classical and quantum contexts.

One could take a further step, citing a close connection between possibility and ignorance: as long as there is ignorance, there are alternative possibilities, and as long as there are alternative possibilities, there is ignorance. But a purely subjective interpretation of probability masks a fundamental difference between the reasons for our ignorance. In a classical world, ignorance results from a practical lack of access to existing facts or data (which could in principle be used to retrodict the past as well as predict the future).%
\footnote{In a cryptodeterministic quantum world, ignorance would result from a theoretical or in-principle lack of access to existing facts or data.}
In our quantum world, ignorance results from the non-existence of facts or data sufficient to  predict the future or retrodict the past.

Why does quantum mechanics reduce us to predicting probabilities? I do not believe that this question can be answered without reference to the constituents of this world---which is what quantum Bayesians refuse to do. (Whatever the constituents may turn out to be, they are not quantum-computational gates.)

The answer I suggest is that the fuzziness (or indefiniteness, or indeterminacy) of all relative position and momenta is an objective feature of the physical world.%
\footnote{I used to consider it superfluous to stress that I am talking about a fuzziness in the world rather than a fuzziness in the mind, but referees's objections against my purported use of ``fuzzy logic'' have taught me otherwise. As I have argued elsewhere~\cite{M2004b,M2006a}, most interpretations of quantum mechanics suffer from a conceptual fuzziness that can be avoided by taking the fuzziness out of the mind and placing it into the world.}
The question then is: how does one define and quantify a fuzzy observable? And the answer is: by assigning probabilities to the possible outcomes of a measurement.  To be precise, the proper way to define and quantify a fuzzy observable is to assign probabilities to the possible outcomes of \textit{unperformed} measurements. 

Suppose we have measured an observable $Q$ and  obtained the outcome~$q$. At the time of the measurement, the value of~$Q$ is~$q$. What can we say about the value, at this time, of an observable $Q'$ that is incompatible with~$Q$? We can say that it is fuzzy. And how do we describe its fuzziness? We describe it by assigning probabilities to the possible outcomes of an unperformed measurement of~$Q'$ on the basis of the actual outcome of the measurement of~$Q$. 

\section{Objective probability}
\label{sec:objective}Contrary to the caveats of CFS, there are several good reasons for thinking of quantum-mechanical probabilities as objective:%
\footnote{I agree, though, that there are also \textit{bad} reasons, e.g., the statistical interpretation of quantum mechanics, the frequentist interpretation of probability, or the propensity interpretation of (quantum-mechanical) probability assignments.}
\bi
\item They are assigned on the basis of (i)~objective, value-indicating events or states of affairs and (ii)~objective physical laws.
\item They do not reduce to Bayesian degrees of belief. (As said, the stability of atomic hydrogen rests on the objective fuzziness of its internal relative position and momentum, not on anyone's belief about the values of these observables.)
\item They play an essential role in the description of physics reality.
\item In particular, they are needed to define and quantify the objective fuzziness of observables.
\ei
I take this opportunity to address an argument against what Fuchs and Schack~\cite{FuchsShack2004} have called ``the folly of trying to have two kinds of probabilities in quantum mechanics''--- one subjective and one objective.

The general state of a spin-$\frac12$ particle can be written in terms of the Pauli matrices and the unit operator~$I$ as
\begin{equation}
\rho=\frac12(I+{\mathbf S}\cdot\sigma),
\end{equation}
where $|{\mathbf S}|\leq1$. If $|{\mathbf S}|<1$, there is an infinite number of decompositions of $\rho$ having the form
\begin{equation}
\rho=\sum_j p_j\,\ketbra{{\mathbf n}_j}{{\mathbf n}_j}\,,
\end{equation}
where $|{\mathbf n}_j|=1$ and $\{p_j\}$ is a probability distribution. Fuchs and Schack appear to think that we are to regard the probabilities $p_j$---which in one decomposition may equal $\frac34$ and $\frac14$, respectively, and in another may both equal~$\frac12$---``as subjective expressions of ignorance about which eigenstate is the `true' state of the particle.'' This is not correct. The probabilities $p_j$ are ignorance probabilities if and only if a measurement with the possible outcomes $\ketbra{{\mathbf n}_j}{{\mathbf n}_j}$ has been made but its outcome is not taken into account. In this case the decomposition of~$\rho$ is uniquely determined by this measurement, and there is no ambiguity whatever about which probabilities are grounded in mere ignorance and which are grounded in objective fuzziness.

On the other hand, if no measurement is made, then $\rho$ itself is the ``true'' state of the particle (that is, the most informative probability algorithm), so the probabilities~$p_j$ cannot be ``subjective expressions of ignorance about which eigenstate is the `true' state of the particle.'' The conclusion of Fuchs and Schack---that ``if a density operator is even partially a reflection of one's state of belief, the multiplicity of ensemble decompositions means that a pure state must also be a state of belief''---thus falls apart.

\section{In search of the ultimate constituent(s)}
\label{sec:UR}The conclusion of Sec.~\ref{sec:macroworld} was that the properties of the microworld supervene on the properties of the macroworld. This means that we cannot think of particles, atoms, and such as constituents of the macroworld. Then what is it that constitutes the macroworld? And first of all, what does it \textit{mean} to constitute something? 

We are in the habit of thinking that small things constitutes large things, and that wholes are made up of distinct individual parts. Quantum mechanics teaches us---at any rate, it ought to have taught us by now---that this kind of bottom-up modeling has passed its expiry date. The conclusion of Sec.~\ref{sec:stdistinctions} was that if we imagine the world partitioned into smaller and smaller spatial regions, there comes a point beyond which these regions, or the corresponding distinctions, no longer correspond to anything in the actual world. By the same token, if we keep dividing a material object, its so-called ``constituents'' lose their individuality, as the following will show.

To begin with, if the properties of the microworld supervene on the properties of the macroworld, then what are the things we call ``particles''? In the first place, they are correlations between ``detector clicks''.%
\footnote{This means, among many other things, that particles do not \textit{cause} detector clicks. An $\alpha$-detector, for example, does not click because an $\alpha$~particle has entered its sensitive region. Rather, the click of an $\alpha$~detector is the reason why an $\alpha$~particle is in the detector's sensitive region.~\cite{UlfBohr2001,M2002}} 
Suppose that we perform a series of position measurements, and that every position measurement yields exactly one outcome (i.e., each time exactly one detector clicks). Then we have a conservation law, and we are entitled to infer the existence of an entity~$O$ which persists through time, to think of the clicks given off by the detectors as indicating the successive positions of this entity, to think of the behavior of the detectors as position measurements, and to think of the detectors as detectors.

If each time exactly two detectors click, we are entitled to infer the existence of two persistent entities. Or are we? If there isn't another conservation law providing the entities with persistent identity tags, then the question of which is which---Which of the particles detected at~$t_1$ is identical with which of the particles detected at~$t_2$?---has no answer. In other words, the question is meaningless.

Here as elsewhere, the challenge is to learn to think in ways that do not lead to meaningless questions. What we have in this case is a \textit{single} entity with the property of being in \textit{two} places whenever we check---not a  system ``made up'' of two things but one thing with the property of being in \textit{two} places every time a position measurement is performed. If we choose this way of thinking, then the meaningless question ``which is which?'' can no longer be asked.

This conclusion can also be reached by other routes. Consider, for instance, a particle that lacks internal relations. What is it ``in itself,'' out of relation to its external relations? The answer is: nothing, except possibly a substance without properties.%
\footnote{Whereas a property is that in the world to which a logical or grammatical predicate can refer, a substance is that in the world to which only a logical or grammatical subject can refer.}
The reason this is so is that the properties of particles are either kinematical relations, such as positions or momenta, or parameters characterizing dynamical relations, such as the various kinds of charge, or they have objective significance independent of conventions only as dimensionless ratios (e.g., mass ratios).%
\footnote{Spin is both relational (inasmuch as its components are defined relative to a reference frame) and characteristic of dynamical relations (inasmuch as it affects a particle's momentum probability distribution in the presence of an electric current). The use of ``dynamical'' is not an endorsement of the notion of quantum state evolution but a reference to the dependence of probabilities on the times of measurements.}

But according to the philosophical principle known as ``the identity of indiscernibles,'' $A$~and $B$ are one and the same thing just in case there is no difference between $A$ and~$B$. Not only is there no difference between two particles lacking internal relations considered ``in themselves,'' but nothing corresponds to the distinction we make between {\it this\/} particle and {\it that\/} particle \textit{over and above} the distinction between {\it this\/} property and {\it that\/} property. What follows from this is the \textit{numerical identity} of all particles lacking internal relations, when each is considered by itself. If we think of particles lacking internal relations as the ``ultimate constituents of matter,'' then there is a clear sense in which the actual number of ``ultimate constituents of matter'' equals~1.

The bottom line: a quantum system is always \textit{one}. The number of its so-called ``constituent'' particles is just one of its measurable properties (which, in a relativistic setting, can come out different every time we check). Hence if I permit myself to think of the entire physical world as a quantum system and to ask about its constituents, I find that it has just one---a single intrinsically propertyless substance. In addition I am now in a position to account for the origin of both matter and space: by entering into spatial relations with itself, this single intrinsically propertyless substance gives rise to space, conceived of as the totality of spatial relations, and it gives rise to matter, conceived of as the corresponding apparent multitude of relata---``apparent'' because the relations are \textit{self}-relations.

\section{Understanding the supervenience of ``micro'' on ``macro''}
\label{sec:manifestation}The conclusions we have reached are supported by a time-honored ontology: ultimately there exists a One Being, and the world is its manifestation. The question, then, is: how does this One Being manifest the world, or manifest itself as the world? And quantum mechanics suggests an answer. The ``microworld'' is neither a world nor a part of any world but instead is \textit{instrumental} in the manifestation of the world (the macroworld, to be precise). Quantum mechanics affords us a glimpse ``behind'' the manifested world at formless particles, non-visualizable atoms, and partly visualizable molecules~\cite{M2006a,M2008}, which, instead of being the world's constituent parts or structures, are instrumental in its manifestation. But---and this is the punch line---we cannot describe what lies ``behind'' the manifested world except in terms of the finished product---the manifested world.%
\footnote{Perhaps Bohr himself didn't realize how right he was when he insisted 
that, out of relation to experimental arrangements, the properties of quantum systems are undefined.~\cite{Petersen1968,Jammer1974}}

But why cannot we describe the manifestation of the world except in terms of the manifested world? The reason is that the spatial distinctions we make---such as the distinction between inside~$R$ and outside~$R$---are warranted only if, and only to the extent that, they are physically realized (made real), for instance by~$R$'s being the sensitive region of a detector. To be able to say truthfully that a particle is inside a region~$R$, we need a detector, and this not merely in order that the particle's presence in~$R$ can be indicated but, in the first place, in order that $R$ be real and the property of being in $R$ be available for attribution to the particle. 

Furthermore, in order that $R$ be real, it must be definable in terms of macroscopic positions. By the same token, in order that the property of \textit{being in}~$R$ be attributable to the particle, it must be indicated by an unpredictable transition in the value of an otherwise deterministically evolving macroscopic position---that of the proverbial pointer needle. And in order that the property of being in~$R$ \textit{at the time}~$t$ be attributable to the particle, $t$~too must be indicated by a macroscopic position. All of this is entailed by the identification of the macroworld as the direct referent of ``reality'' or of ``the (physical) world.''

\section{Conclusion}
\label{sec:conclusion}The purpose of this paper was twofold: (i)~to offer a critique of the Bayesian interpretation of quantum mechanics with particular focus on a paper by CFS (which contains a critique of the ``objective preparations view'' or OPV), and to carry the discussion beyond the hardened positions of the Bayesians and the proponents of the OPV. Several claims made by CFS were rebutted, including the claim that different pure states may legitimately be assigned to the same system at the same time, and the claim that the quantum nature of a preparation device cannot legitimately be ignored. 

If a quantum state is more than a probability algorithm, it is not testably so. It is therefore only prudent to view the time dependence of a quantum state as a dependence on the time of the measurement to whose possible outcomes the quantum state serves to assign probabilities. For both Bayesians and proponents of the OPV, on the other hand, the time dependence of a quantum state is the continuous dependence on time of an evolving state of some kind. This makes it impossible to recognize the full implications of the one and only testable feature of quantum mechanics---the probability distributions it yields. For, as shown, continuous state evolution entails a completely differentiated time and therefore a completely different spacetime, whereas an analysis of quantum-mechanical probability distributions yields the opposite: the spatiotemporal differentiation of the physical world does not go ``all the way down.''

Why are measurements essential ingredients of all standard formulations of quantum mechanics? While Bayesians keep the measurement problem safely locked away in ``a single black-boxed piece of hardware''~\cite{Mermin2006}, most others try to gloss it over or explain it away. To be able to understand the special status of measurements, one needs a clear distinction between what counts as macroscopic and what does not. As shown, it is the incomplete spatiotemporal differentiation of the physical world that makes this distinction possible. This incomplete differentiation is entailed by an objective fuzziness, which is required for the stability of matter, and which explains why the fundamental theoretical framework of physics is a probability calculus. 

Bohr was right: out of relation to measurements, the values of physical observables are undefined. They are undefined because they supervene on value-indicating events or states of affairs. What constitutes the macroworld is not particles, atoms, or such but a Single Being, an Ultimate Reality, about which we know nothing, except that it manifests the macroworld or manifests itself as the macroworld. The microworld is neither a world nor a part of any world but instead is instrumental in the manifestation of the macroworld. Quantum mechanics affords us a glimpse ``behind'' the manifested world, at stages in the process of manifestation, but it does not allow us to describe what lies ``behind'' the manifested world except in terms of the finished product---the manifested world. This is the reason why the properties of the microworld supervene on those of the macroworld. The reason why the manifestation of the world cannot be described except in terms of the manifested world, is simply that without the manifested world there is nothing in whose terms we could describe its manifestation.

\end{document}